\documentclass[10pt, preprint, superscriptaddress]{revtex4-1}

\usepackage{graphicx}
\usepackage{dcolumn}
\usepackage{bm}
\usepackage{indentfirst}
\usepackage[greek,english]{babel}

\usepackage{epsfig}
\usepackage{graphics}
\usepackage{booktabs}

\usepackage[latin1]{inputenc}
\usepackage{setspace}
\usepackage{makeidx}

\usepackage[mathscr]{eucal}
\usepackage{mathrsfs}
\usepackage{latexsym}
\usepackage[usenames,dvipsnames,svgnames,table]{xcolor}
\usepackage{textgreek}
\usepackage{float}

\begin{document}

\title{Deep neural network for X-ray photoelectron spectroscopy data analysis}

\author{G. Drera} 
\email{giovanni.drera@unicatt.it}
\affiliation{I-LAMP and Dipartimento di Matematica e Fisica, Universit\`a Cattolica del Sacro Cuore, Brescia I-25121, Italy}

\author{C.M. Kropf} 
\affiliation{I-LAMP and Dipartimento di Matematica e Fisica, Universit\`a Cattolica del Sacro Cuore, Brescia I-25121, Italy}
\affiliation{Istituto Nazionale di Fisica Nucleare, Sezione di Pavia, via Bassi 6, I-27100 Pavia, Italy}

\author{L. Sangaletti} 
\affiliation{I-LAMP and Dipartimento di Matematica e Fisica, Universit\`a Cattolica del Sacro Cuore, Brescia I-25121, Italy}

\keywords{XPS, Photoelectron Spectroscopy, Deep Convolutional Neural Networks}


\date{\today}

\begin{abstract}
In this work, we characterize the performance of a deep convolutional neural network designed to detect and quantify chemical elements in experimental X-ray photoelectron spectroscopy data. Given the lack of a reliable database in literature, in order to train the neural network we computed a large ($>$100 k) dataset of synthetic spectra, based on randomly generated materials covered with a layer of adventitious carbon. The trained net performs as good as standard methods on a test set of $\approx$ 500 well characterized experimental X-ray photoelectron spectra. Fine details about the net layout, the choice of the loss function and the quality assessment strategies are presented and discussed. Given the synthetic nature of the training set, this approach could be applied to the automatization of any photoelectron spectroscopy system, without the need of experimental reference spectra and with a low computational effort.
\end{abstract}

\maketitle

\section{Introduction}
Deep neural networks (DNNs) are currently state-of-the-art in image recognition applications, and have already been tested for several scientific spectroscopy applications\cite{DNN_VB_spectra,DNN_Raman,DNN_nature,Chatzidakis2019}. In fact, fast machine learning processing will be crucial for high-throughput data analysis, especially for large research experiment facilities such as synchrotron or free-electron lasers\cite{DNN_sepe}, where the large data amount prevents the standard hand processing. In addition to the way faster processing, machine learning methods can, for specific tasks, match or even outperform the accuracy of a human analysis.

\begin{figure}[ht!]
\begin{center}
\includegraphics[width=0.5\textwidth]{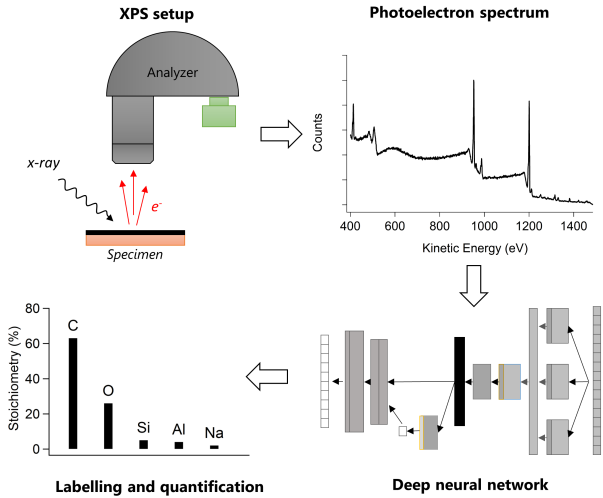}
\caption{Schematics of XPS and of the DNN application to spectra analysis.\label{fig_abstract}}
\end{center}
\end{figure}

X-ray photoelectron spectroscopy (XPS) data represent an ideal application field for deep neural network (DNN) classification methods. In an XPS  experiment \cite{XPS_Fadley} the sample surface is hit by x-rays with specific energy ($h\nu$) from a monochromatic source (schematics are given in Figure \ref{fig_abstract}). If $h\nu$ is larger that the binding energy (BE) of the electrons in the solid, the electron is ejected with a kinetic energy $KE = h\nu - BE - \phi$, where $\phi$ is the work function, ultimately related to the bulk/vacuum discontinuity at the sample surface. This simple relation allows one to collect XPS spectra by measuring the number and the kinetic energy of the photoemitted electrons.

Each element displays characteristic core levels and Auger lines that are then used to identify the elements present in a sample. Elemental identification should be obtained with ease with DNNs, without the complexities required, for example, for image recognition tasks. Given the potential relevance of XPS for material physics and industrial research\cite{STICKLE201950}, this machine learning approach looks appealing for automated analysis.

However, major drawbacks prevents the application of DNNs to XPS data analysis. A neural network training requires a large database of consistent spectra, which should cover all the possible XPS analysis outcomes in a well distributed, random sampling of all chemical elements; up to now, such amount of data cannot be found in literature. The lack of a proper spectra database is due to several characteristic of XPS analysis: the technical complexity (due to the ultra high vacuum requirements), the variety of XPS set-up (different photon sources, analyzers, experimental geometries), the often long spectra acquisition time, the details of sample preparation, and the different spectra range, resolution and noise level. Apart from time and sample constraints, the collection of a universal XPS database suitable for DNNs is unfeasible, since each XPS machine would require a special dataset related to its specific technical details.

Moreover, the XPS quantification and identification process is strongly influenced by the large difference of photoemission cross sections for each chemical element. In fact, the actual detection threshold is different for each element, and is also dependent on the actual total electron count statistics; a larger acquisition time or a higher photon flux allows\cite{Hill_2017_part1} to detect elements in a sample with a sensitivity down to 0.1\%. Moreover, the peculiar superposition of core-levels of different elements significantly affect the XPS element detection capability\cite{Shard_2014}. The stoichiometry evaluation (i.e., the elemental quantification) is also affected by the spectra analysis routine and by the specific choice of sensitivity factors. The practical accuracy for relative elemental quantification is generally considered\cite{Seah_2001} to be 10\%, although much better results can be obtained with a very accurate setup characterization and data treatment\cite{Drera_TransCal}. As a result, most of the quantification work is usually carried out on a single, specific core-level for each element, for which a specific sensitivity factor is known\cite{Seah_reference_book}; these normalization factors are different for each XPS setup and must be supplied by the machine vendor or obtained by accurate investigations on calibration samples. An extensive review of quantitative XPS resolution can be found at this reference\cite{Faradzhev_2017_part2}.

Finally, XPS spectra are often affected by the presence of a surface adventitious carbon contamination layer, due to the high surface sensitivity of the technique. While in some cases this layer can be removed by UHV cleaning techniques, such as Ar$^+$ sputtering or plasma cleaning, in many other cases the surface can not be cleaned without inducing a sample degradation. The carbon contamination leads to an overall lower XPS intensity\cite{evans_contamination}, which is different for each core level because of its dependency on the photoelectron KE.

In this work we show the application of a DNN to the task of identification and quantification of XPS survey spectra. To overcome the lack of a large experimental dataset, we generated a synthetic training set, based on state of the art theory for XPS\cite{Werner_MonteCarlo}. Each training spectrum has been calculated on a randomly generated material, with a random contamination layer on top; every element in the periodic table, from Li to Bi, has been taken into account with identical probability. Each detail of real XPS spectra, such as peak position, width and intensity, inelastic loss backgrounds, chemical shifts, the analyzer transmission function, the signal-to-noise ratio etc., has been carefully simulated according to available XPS databases and theories, in order to produce random synthetic, yet realistic spectra. As for an experimental data reference, we used a set of 535 survey spectra, collected in similar experimental conditions.  The DNN has been specifically designed to produce consistent results without the use of any experimental spectra during the training; for this task, an optimized net layout and a specific loss measurement have been introduced. The DNN has also been trained to ignore the adventitious carbon contribution, in order to produce the pristine material stoichiometry quantification. Due to its design, this approach could be applied to any XPS system, with any photon source, and without the need of a large experimental data set for the DNN training.

\section{Methods}
\subsection{The training set}
We numerically generated a synthetic training set made of 100k survey spectra based on XPS parameter databases and electron scattering theory in the transport approximation\cite{Werner_MonteCarlo}. The spectra KE range was 400-1486 eV on a 2000 point grid, which then will correspond to the size of the input array of the DNN (i.e., the number of features); we restricted the analysis to Al k$_{\alpha}$ source, although this method could be extended to any soft x-ray source, within the limits of database availability and model approximations.

\begin{figure}[ht!]
\begin{center}
\includegraphics[width=0.48\textwidth]{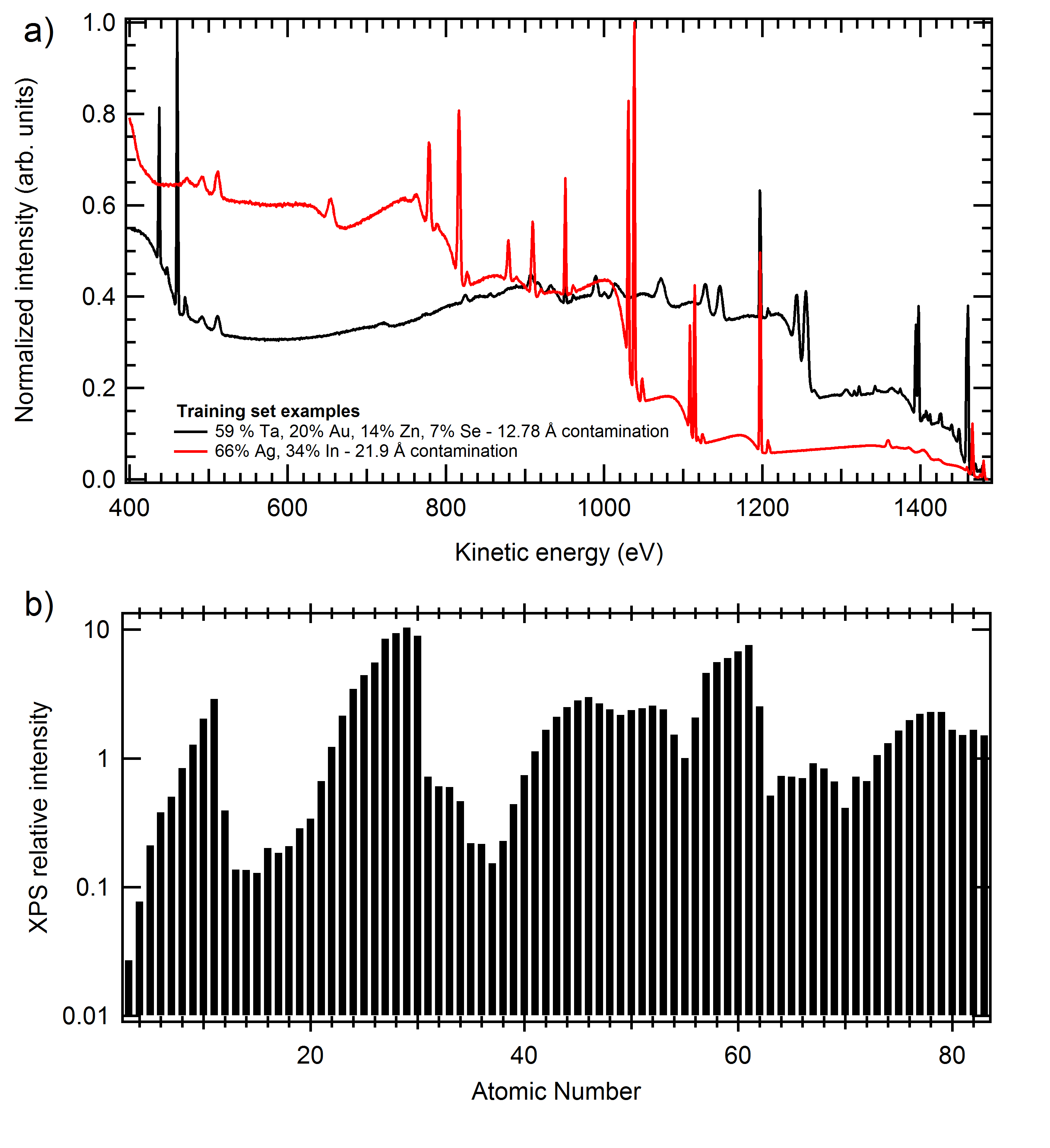}
\caption{a) an excerpt of two simulated survey training spectra; b) maximum XPS intensity of simulated spectra ($T_i$) for each element;  \label{fig_thresh_examples}}
\end{center}
\end{figure}

Given that we only consider the task of overall identification and elemental quantification, we devised a relatively simple two-layer model for the spectra simulation, where a random bulk material is covered by an over-layer of the usual hydrocarbon contamination. In order to make the training set as general as possible, each material is composed by a random number (from 2 to 5) of elements, with variable stoichiometry ratios. We consider random possible combinations of elements in the $[3,81]$ atomic number interval, without any bias towards a specific element or material; while this method is generating spectra for several unphysical materials, it allows for a completely unbiased network training. We assigned to each synthetic material a density, evaluated on the basis of elemental densities. Although this is a rather approximate approach, it allows for a more accurate evaluation of peak intensity in very dense materials, such as low-Z elements diluted in high-Z compounds. The contamination layer density has been set to 1.56 g$/$cm$^{-3}$, as an average of several similar organic compounds; the environmental contamination layer should thus be considered as an effective layer, whose thickness was randomly chosen in the $[0-40]$ {\AA} interval.

For each element we considered databases entries for all the core-levels\cite{YehLindau,TNY_cross} and Auger\cite{Hubbel_auger_fluorescence,Coghlan_auger} structures, both for cross-sections and native peak widths. Peak positions are also randomly shifted by considering the largest chemical shifts found in available databases\cite{NIST_xps_database}; this shift can reach up to 10 eV, for instance for sulfur core levels. In order to predict the actual peak intensity we performed full Monte-Carlo simulations, as described by Werner\cite{Werner_MonteCarlo}, including both the electron inelastic and transport mean free path (IMFP and TMFP). IMFP has been calculated with the usual TPP2M formula\cite{TPP2M}, while for TMFP we used the interpolation method of Jablonsky\cite{TMFP}. This method allows for the prediction of peak intensity and of the peak inelastic background, which has been simulated on the basis of Tougaard differential inverse inelastic mean free path (DIIMFP) formula\cite{tougaard_diimfp}. Each peak has been simulated with a Voigt peak, and the background has been evaluated through subsequent convolutions with the DIIMFP function. The full details of the actual XPS setup have been considered, including the analyzer acceptance angle. Such approach has been used previously to accurately characterize the transmission function of electron analyzers from survey spectra quantification\cite{Drera_TransCal}.

For the contamination layer, based on experimental results on adventitious carbon contamination\cite{evans_contamination}, we considered a carbon rich layer (5:1 carbon to oxygen ratio), with an additional 10\% noise for the intensity and a random shift of maximum 0.5 eV for the peak position. Such shift should mimic the effect of small peak drifts due to charging or to different analyzer-to-sample work functions.

In order to make comparison with the raw experimental spectra, we performed several final refinement steps. Synthetic spectra are convoluted with a gaussian peak in order to reproduce the experimental resolution; we introduced the XPS satellites for the non-monochromatic Al k$_{\alpha}$ source(for core-level photoelectron features only); we multiplied the spectra with the proper analyzer transmission function, which has been characterized properly for our XPS system\cite{Drera_TransCal}; finally, we added a small gaussian relative noise (0.3 \%) and normalized the data in the $[0,1]$ intensity range. Some example of the training set spectra are given in Figure \ref{fig_thresh_examples}-a.

The generation of each synthetic spectrum required an average of two minutes computational time on a single-core desktop machines. Consequently, the training set production was in fact the most computational intensive part for this work, and was carried out in parallel fashion on several computers. The relatively long time required for a single spectrum calculation and the large number of parameters renders the application of this spectra prediction method as a direct data fitting procedure impracticable. Instead, after the DNN training, any subsequent spectra evaluation is then extremely fast and direct.

The $[0,1]$ intensity normalization constraint, which is typical for the input features of trainable DNNs, introduces some additional difficulties for the quantification task. The XPS total intensity, measured with a fixed X-ray flux and accumulation time, can vary by up to two order of magnitude because of the element cross sections. Accordingly, the amount of time required to decrease the noise level of XPS spectra is also variable. By fixing the intensity range, we are then losing information which could be in principle useful for the quantification process. However, in standard XPS practice, the X-ray flux could be different in each experiment, as well as the accumulation time; hence the choice to normalize the training and experimental spectra to the same scale. For the same reason we also decided to fix the signal-to-noise ratio (SNR) to 0.5\% of the $[0,1]$ intensity range (i.e., each spectra shows exactly the same SNR), which corresponds to the lab practice to tune the total accumulation time in order to reach a reasonably clean spectrum.

Two different set of labels have been considered for the DNN output. The most straightforward one is the choice of an 81 numbers array $\overline{q}_i$ $i=1,\dots,81$, which directly represent the relative elemental quantification from $Z = 3$ (lithium) to $Z = 83$ (bismuth). However, this approach is suboptimal because the relative elemental quantification does not directly correspond to the relative contribution of each chemical species to the corresponding spectrum intensity, which confuses the DNN. More precisely, due to the different photoelectron cross sections, each element displays different XPS intensities for the same relative quantification. For instance, in a compound made by Li and Cu in a 1:1 ratio, 99\% of the XPS spectral weight is related to Cu, making the lithium detection nearly infeasible without a very large data statistics. In Figure \ref{fig_thresh_examples}-b, we show as a reference for the relative XPS intensity the calculated spectra maxima of pure elements without contamination (labeled $T_i$). Therefore, instead of the relative elemental quantification, we used as labels for the classification the normalized quantification intensity defined as

\begin{equation}
\overline{y}_i = \overline{q}_i  T_i  (\sum_i \overline{q}_i T_i  )^{-1} \label{eq_normalization}
\end{equation}
where now $\overline{y}_i$ is the contribution of the element $i$ to the total intensity spectrum. Here and throughout this paper variables with overline denote true labels and variables without overline denote the network outputs.

\subsection{The experimental data}
The experimental dataset is composed of 534 survey spectra  collected in the Surface Science and Spectroscopy laboratory of the Università Cattolica in Brescia with a VG-Scienta R3000 spectrometer and the non-monochromated K$_\alpha$ line of a PsP dual-anode X-ray source. The dataset contains several classes of materials: many inorganic oxides, binary and elemental semiconductors, carbon based nanostructures and heterostrutures. Of the 81 element used in the generation of the synthetic training set, 36 are actually composing the materials in the experimental dataset.
The only data correction which has been applied to the experimental spectra before testing them with the trained DNN is the alignment of the energy scale, carried out with reference peaks energy (Ag, Au or adventitious carbon contamination). None of the experimental data has been used as a part of the training set, and none of the synthetic training set data has been modified to fit the experimental results.

In order to obtain the labels $\overline{q}_i$ of the experimental spectra we first performed a standard quantification by using the peak area of all the detectable elements. We then removed the contribution of surface carbon contamination from the experimental labels, whenever allowed by the additional info about each specimen. Such evaluation can be problematic even for a human user, especially for heavily contaminated carbon-based materials. With this method, the obtained labels $\overline{q}_i$ have a relative error of $\sim 10\%$ on the experimental dataset.

\subsection{Deep neural network layout}

\begin{figure}[ht!]
\begin{center}
\includegraphics[width=0.5\textwidth]{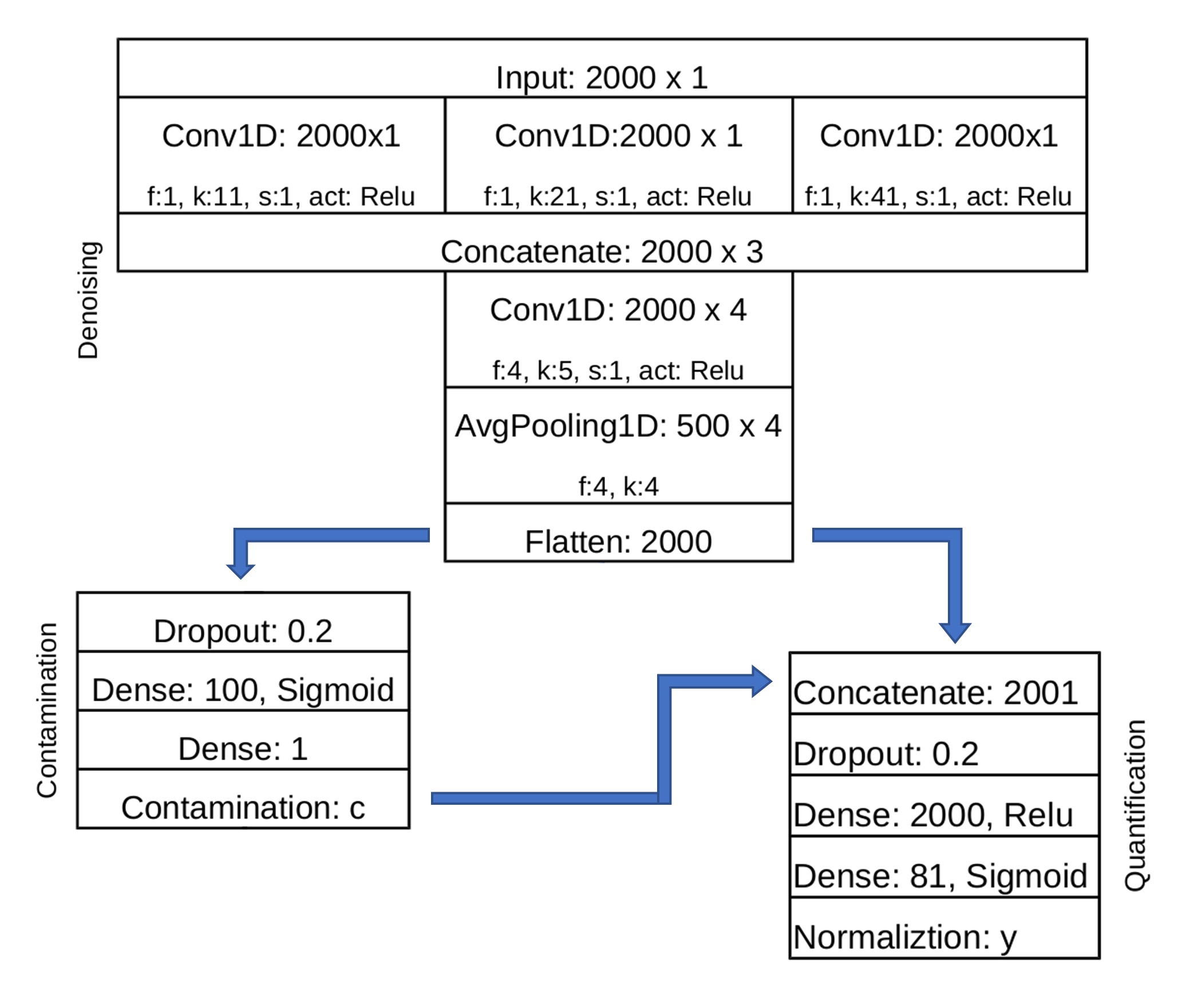}
\caption{Deep neural network layout for XPS data identification; for each convolution layer, $f$ is the filter number, $k$ is the convolution kernel size, $s$ is the stride length. Total number of trainable parameters is $4\,366\,410$.\label{fig_DNN}}
\end{center}
\end{figure}

We tested several geometries for our neural network and, although we tested some purely fully connected layer and deep convoluted networks, the best results were obtained for the hybrid geometry shown in Figure \ref{fig_DNN}, which also identifies the level of carbon contamination. The network takes as input the 2000 spectral points $x_j$ ($j=1,\ldots,2000$) and produces two outputs, the normalized contamination level $c \in [0,1]$ (equivalent to $[0-40]\AA$) and the normalized intensity $y_i$ ($i=1,\ldots,81$) of the 81 element, c.f. Eq.~(\ref{eq_normalization}). In order to obtain the actual quantification $q_i$ ($i = 1 ,\ldots 81$) of the elements, we post-process the learned outputs of the network

\begin{equation}\label{eq:postProcessing}
	q_i = y_i / T_i \left(\sum_i y_i\right)^{-1}.
\end{equation}

Our best network is composed of three stages. The first is made of convolutions and essentially plays the role of a noise filter and feature extraction. The second is used to identify the level of carbon contamination, and the third is used for normalized intensity quantification.

Concretely, the first stage is made of an initial multilevel convolution sub-net modeled after the inception module of incnet \cite{Szegedy2015}, with three parallel 1D convolutions with one kernel each, of size $11,21,41$, respectively, and stride length $1$. The resulting data are then concatenated, pooled (with an average pooling kernel size 4, stride length 4) and flattened to obtain again 2000 data points.
The second stage is composed of two fully connected layers with $100$ and $1$ neurons, both with (logistic) sigmoid activation, is used to identify the level of carbon contamination $c$ (loss: $L_2$-norm).
The third stage produces the normalized intensities $y_i$. For this, the output $c$ is concatenated with the 2000 elements of the first convolution stage. The 2001 elements are then used as input for the last fully connected classification layers of 2000 neurons with a rectified linear unit (RELU) activation and 81 neurons with sigmoid activation, respectively. A final layer is used to normalize the outputs so that $\sum_i y_i = 1$. For the full network, the total number of trainable parameters is about 4 million.

The optimal net hyper-parameters, such as the convolution kernel sizes, the convolution number, the fully-connected layer sizes and the optimization algorithm have been tuned with hyperparameter optimization routines. One of the major difficulties in the hyper-parameter optimization was the choice of the proper activation functions and the associated loss function for the quantification stage (81 output neurons). From a machine learning perspective, we deal with a multi-label supervised learning task, which points towards using a sigmoid output on each of the output neurons together with a binary-cross-entropy loss. However, contrarily to the standard definition of multi-label classification tasks \cite{scikit-learn}, the labels are not independent, but must sum up to one as we consider \emph{relative} concentrations of elements. Thus, one could use a soft-max activation with categorical cross-entropy. However, this produces unsatisfying results from a physics perspective, because it cannot deal well with samples with several roughly equally present elements. For practical purposes, one would desire a network that has a very high accuracy for large relative concentrations, let's say $\geq 10\%$. In order to achieve this, we used the combination of sigmoid activation functions, a non-trainable normalization layer (as shown in Figure ~\ref{fig_DNN}) and a custom loss function
\begin{equation}
	L(y,\overline{y}) = \sum_{i=1}^{81} y_i^2 (y_i - \overline{y}_i)^2
\end{equation}
where $y_i$ is the network output and $\overline{y}_i$ the target values. This loss function, which multiplies the standard L$^2$ norm by the net results squared, is then larger for large output values; for this reason we termed it 'high-pass filter' loss. The combination of this net layout, ADAM optimizer and the tuned loss function allowed for a robust training.

\begin{figure}[ht!]
\begin{center}
\includegraphics[width=0.45\textwidth]{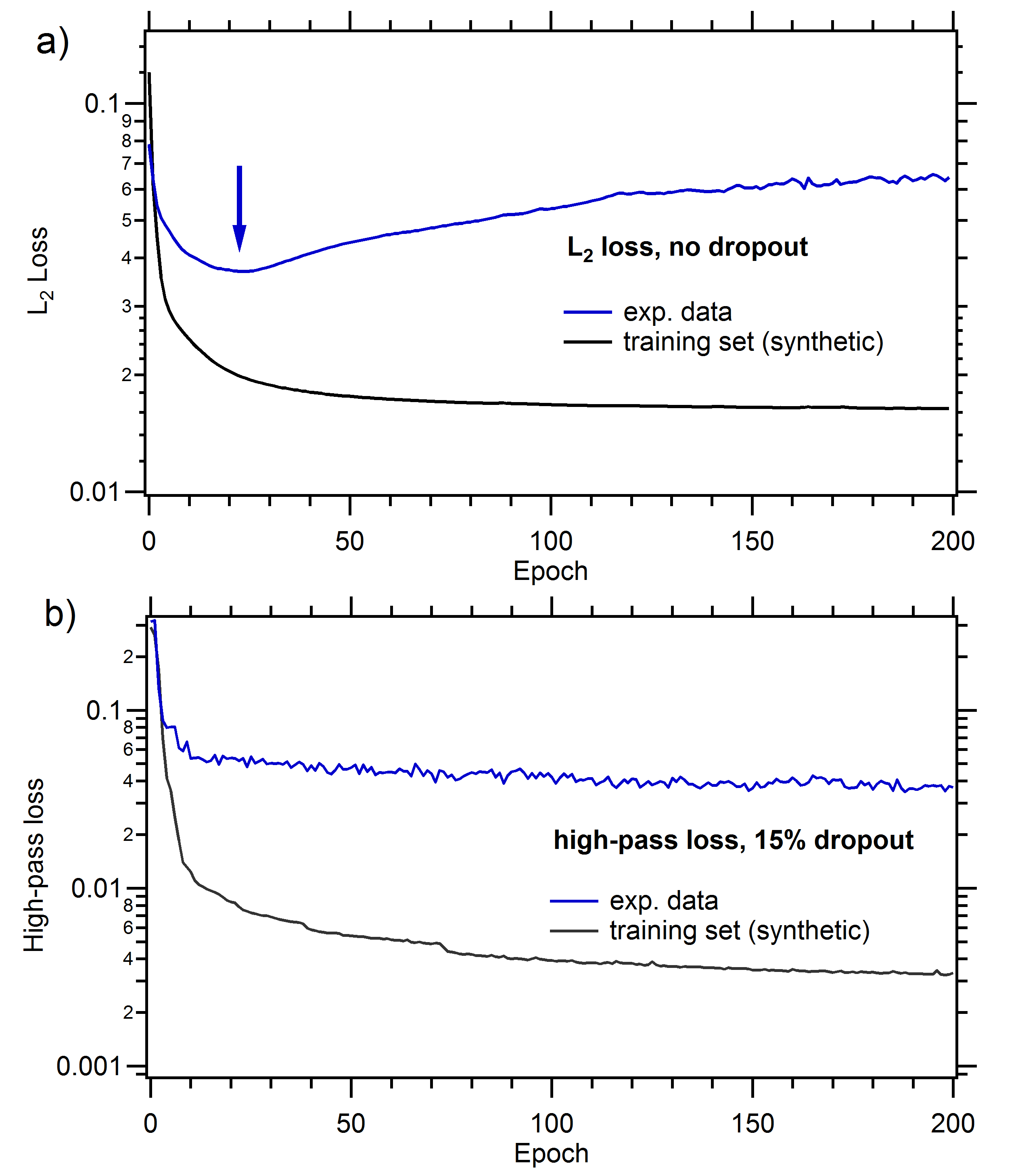}
\caption{Loss function convergence during training for a) no dropout and standard L$^2$ norm and b) 15\% dropout before the classification layer and high-pass norm. \label{fig_training_loss}}
\end{center}
\end{figure}

\begin{figure*}[ht!]
\begin{center}
\includegraphics[width=0.95\textwidth]{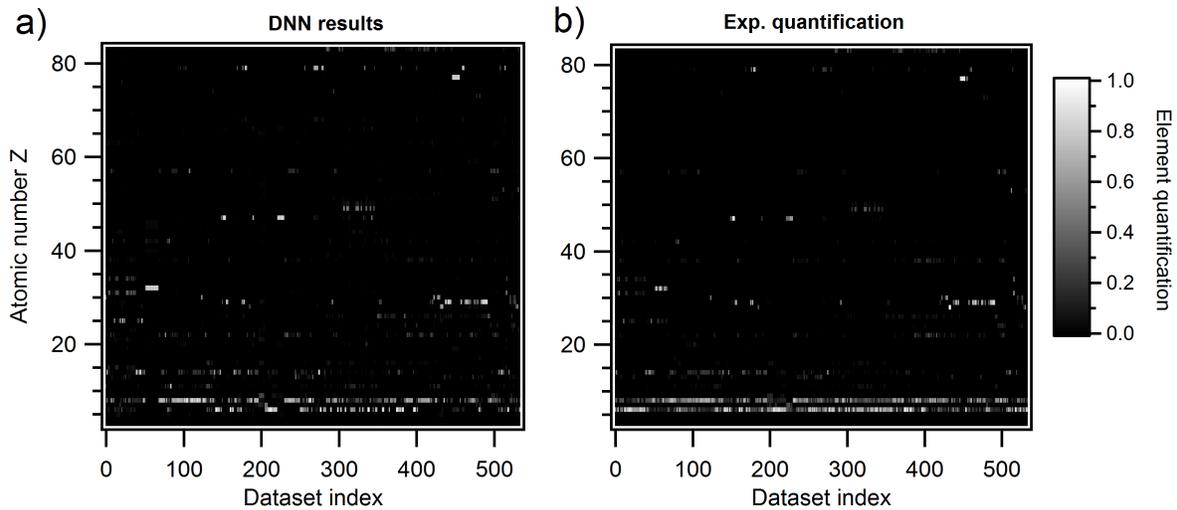}
\caption{DNN quantification results (a) compared to actual XPS quantification (b).  \label{fig_quantification_spectra}}
\end{center}
\end{figure*}

\section{Results and discussion}

The evolution of the loss function for the DNN training is given in Figure \ref{fig_training_loss}.
Without the introduction of a regularization methods, after few epochs the training algorithm begins to overfit the synthetic data, leading to a minimum in the experimental data loss (blue arrow in Figure \ref{fig_training_loss}-a). It should be pointed out that overfitting is observed on the experimental data only, and not on subsets of synthetic spectra (not used during the training); the dataset size is thus adequately large, and has been computed with an sufficient amount of randomness in the initial parameters.

In order to avoid overfitting and to achieve a better parameter convergence we introduced a dropout before the final classification layers (c.f. Figure \ref{fig_DNN}): the training algorithm was then forced to ignore a random portion of the net connections at a specific location, with a probability $p$. With this method we routinely achieved a smooth convergence of both the training set and the experimental data (see Figure \ref{fig_training_loss}-b).
As a side effect, the introduction of a dropout layer led to a slightly slower learning process; we estimated that a training of nearly 200 epoch is enough to achieve good and consistent quantification and identification performances. The optimal range for the dropout probability, estimated by the hyperparameter study, is $0.1<p<0.2$; within this interval the dropout is large enough to avoid overfitting, and small enough to prevent excessive randomness in the quantification performances.

The quantification results $q_i$ obtained using the post-processing Eq.~(\ref{eq:postProcessing}) from the predicted labels $y_i$ of a well trained DNN applied to the experimental data are given in Figure \ref{fig_quantification_spectra}. The DNN quantification results $q_i$ (Figure \ref{fig_quantification_spectra}-a) are contrasted to the experimental quantifications before the removal of oxygen and carbon from the adventitious contamination layer (Figure \ref{fig_quantification_spectra}-b); data are shown as matrices where each column corresponds to a spectrum and each row correspond to a specific element. The overall correspondence between the DNN predictions and the actual quantifications is remarkable, with a complete absence of wrong element detection with relative high stoichiometry. Factoring out the presence of adventitious contamination in the experimental data, the RMS between $q_i$ and $\overline{q}_i$ is equal to 3.8 \%. However, this value can only poorly asses the DDN performances due to the inherent uncertainty in the experimental labels $\overline{q}_i$.

\begin{figure}[ht!]
\begin{center}
\includegraphics[width=0.46\textwidth]{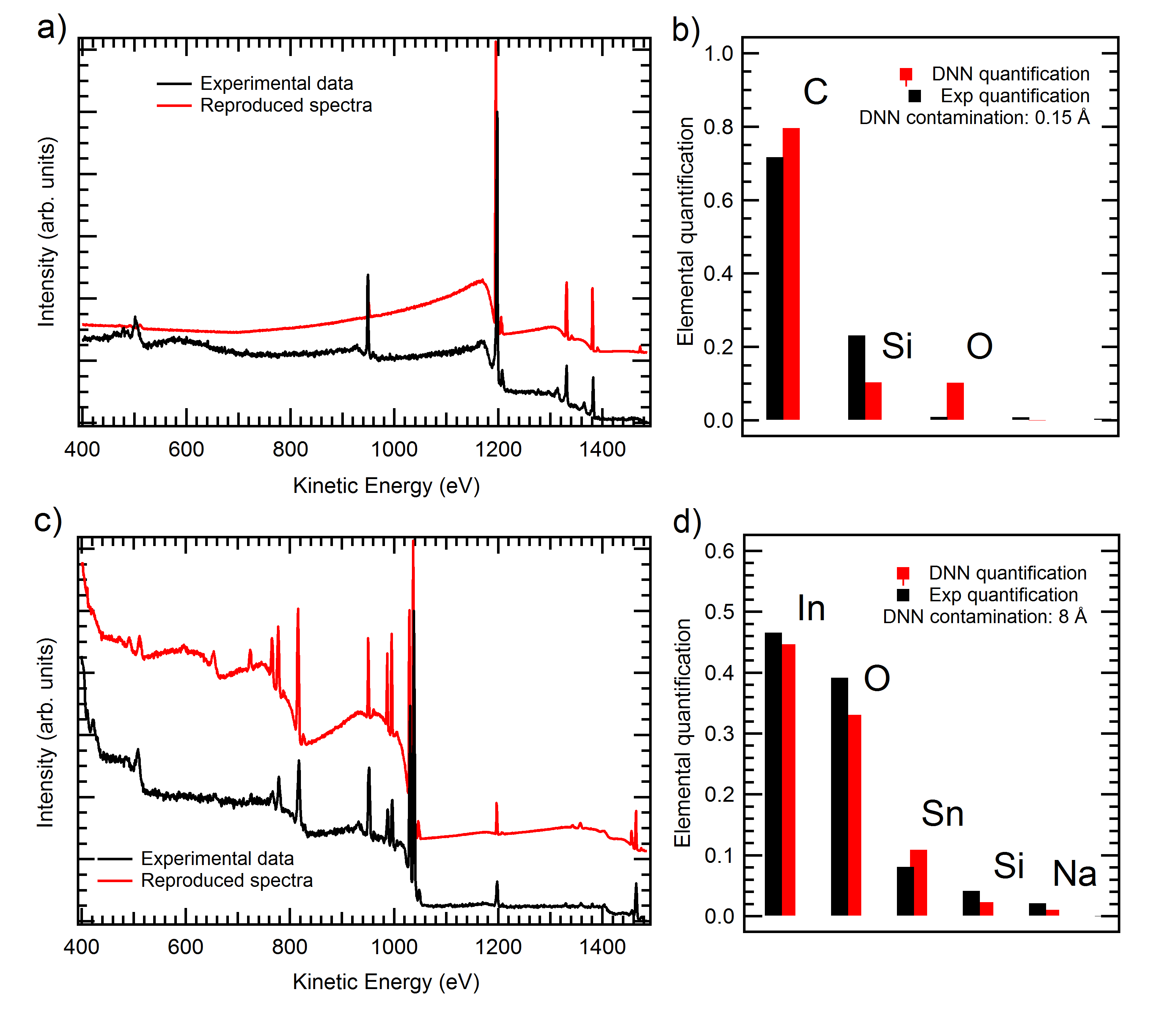}
\caption{Application examples on two experimental sets; spectra (a) and quantification $q_i$ (b) for carbon nanotubes on Si; spectra (c) and quantification $q_i$ (d) for indium-tin oxide (ITO) deposited on Si. For the ITO case, the adventitious carbon contamination was factored out from the experimental quantification. \label{fig_examples}}
\end{center}
\end{figure}

Two examples of actual data quantification are given in Figure \ref{fig_examples}. For carbon nanotubes deposited on Si (Figure \ref{fig_examples}-a), the DNN correctly assign the carbon peak to the actual material and not to the adventitious contamination (Figure \ref{fig_examples}-b), which is nearly negligible; in indium-tin oxide (Figure \ref{fig_examples}-c,d), the DNN correctly identifies all the main elements and assigns all the carbon content to the actual adventitious contamination. It is also possible to use the DNN stoichiometry and contamination predictions to actually reproduce synthetic spectra (red traces, Figure \ref{fig_examples}-a,c) which are in a nice agreement with the experimental one, within the 10\% accuracy limit of the hand-made quantification methods.

\begin{figure}[ht!]
\begin{center}
\includegraphics[width=0.46\textwidth]{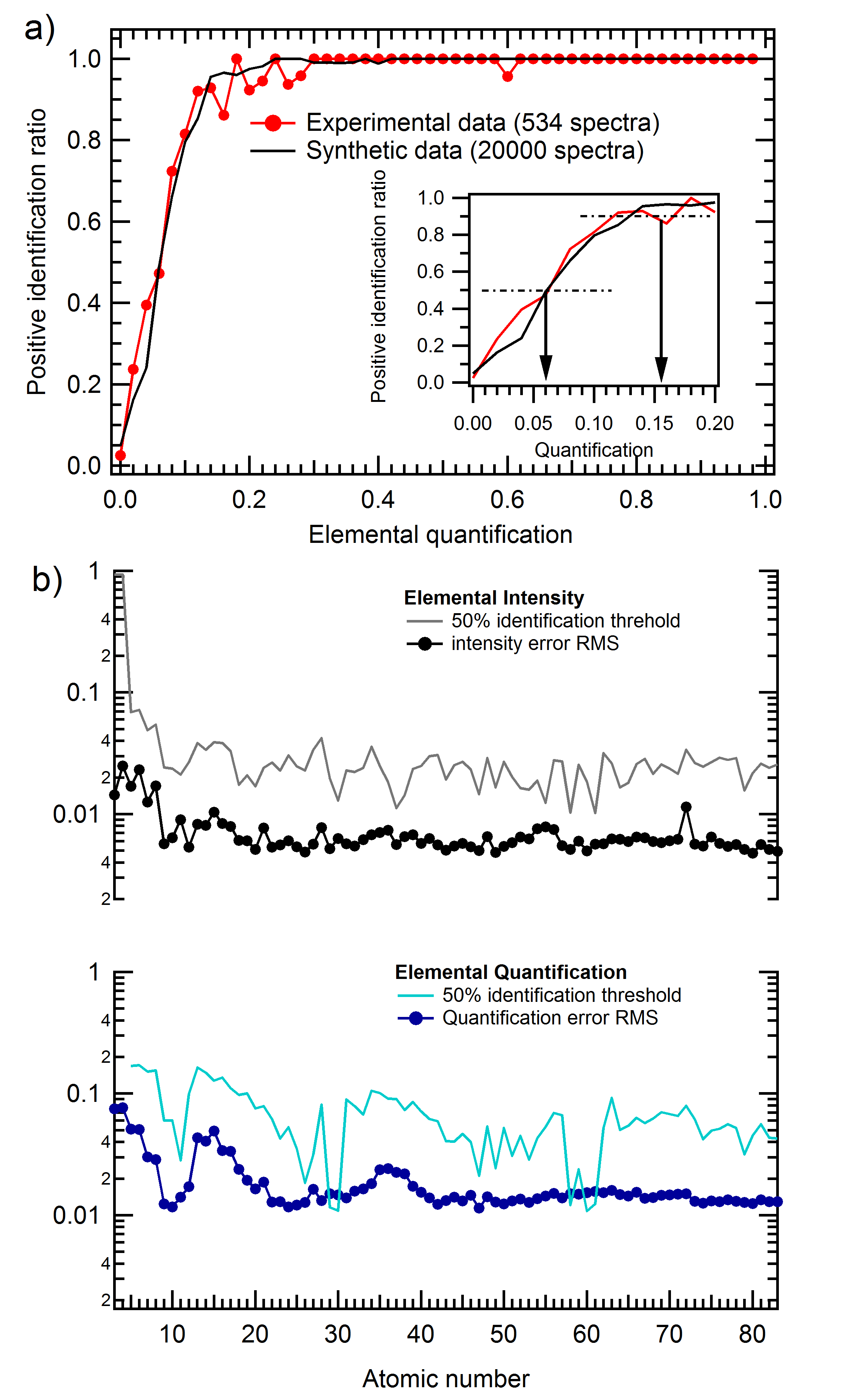}
\caption{a), positive identification ratio of the DNN vs the quantification output, calculated for the experimental data (red) and a synthetic training set (black); b) element specific identification thresholds for the intensity $y_i$ and the quantification $q_i$, calculated for the training test set. \label{fig_histogram_selective}}
\end{center}
\end{figure}

In order to further asses the detection accuracy of the DNN we considered the identification ratio metric as shown in Figure\ref{fig_histogram_selective}-a. We first divided the interval of values for the $q_i$ ([0,1]) into 50 bins of equal length $0.02$. Then, for each bin, we counted the ratio of elements predicted by the DNN that have a $q_i$ within the bin that also have a label $\overline{q}_i>0$. This was done for the whole experimental data set (red dot line) and for a set of 20000 synthetic spectra not used during the training (black line). Since the DNN uses a sigmoid function on the output neurons (c.f. Figure \ref{fig_DNN}), the predicted quantifications $q_i$ for all the elements is always a non-zero.

For a well-working network, we expect most of the wrong predictions to be related to small output (either $q_i$ or $y_i$), while positive identification should be related to high output values. With a perfect identification, the graph of Figure \ref{fig_histogram_selective}-a should then be a Heaviside-like step function, whose integral $I_R$ in the $[0-1]$ output interval would be exactly one. We found that a good elemental identification, regardless of the RMS on the quantification accuracy discussed above, can be found when the integral of this function is above 0.90; with our DNN we routinely achieve a $I_{exp} \approx 0.93$, such as for the red dotted curve shown in Figure \ref{fig_histogram_selective}-a. For the synthetic dataset (black line in Figure\ref{fig_histogram_selective}-a), the integral is $I_{synth}=0.95$, which is very close to $I_{exp}$.

The identification graph can now be used to estimate an accuracy threshold for the DNN results, pointed out by black arrows in the Figure \ref{fig_histogram_selective}-a inset; an average 50\% probability of a correct element identification is found for DNN quantification $q_i$ results above 0.06, while the 90\% threshold is reached roughly at 0.15. When the DNN output $q_i$ for a specific element is larger than 0.2, the identification is nearly always correct. We tracked the value of $I_{exp}$ over several different training runs and obtained and average of $0.92 \pm 0.01$ level after 200 epochs.

Although these thresholds could be used as general rule-of-thumb, we expect to have different accuracies for different elements as a result of the different photoelectron cross sections. However, the limited size of our experimental dataset did not allow for a precise elemental-selective accuracy study. Instead, we computed the positive identification curves over the synthetic test training set each element individually. This was done for both the final quantification ($q_i$) and the network output intensity ($y_i$). The corresponding 50\% identification threshold and RMS of the net output with respect to the exact labels are shown in Figure\ref{fig_histogram_selective}-b. The net performances for elemental intensities (black and grey graphs) is nearly constant for all the elements, with the exception of Li and Be, due to their very low photoelectron cross sections; the average value for the 50\% threshold is close to 0.03, i.e. the the DNN is able to detect an element contributing to 3\% of the total spectra intensity, with an average error of about 0.7\%. The quantification on the contrary (blue and cyan traces) is strongly dependent of the atomic number, with a detection threshold ranging from 1\% to nearly 20\%; accordingly, the absolute accuracy of quantification is also varying from a base value of 1.5\% to 9\%. Note that for both identification and quantification, the lower limits are probably dependent on the amount of random noise added to the synthetic training set.

\begin{figure}[ht!]
\begin{center}
\includegraphics[width=0.48\textwidth]{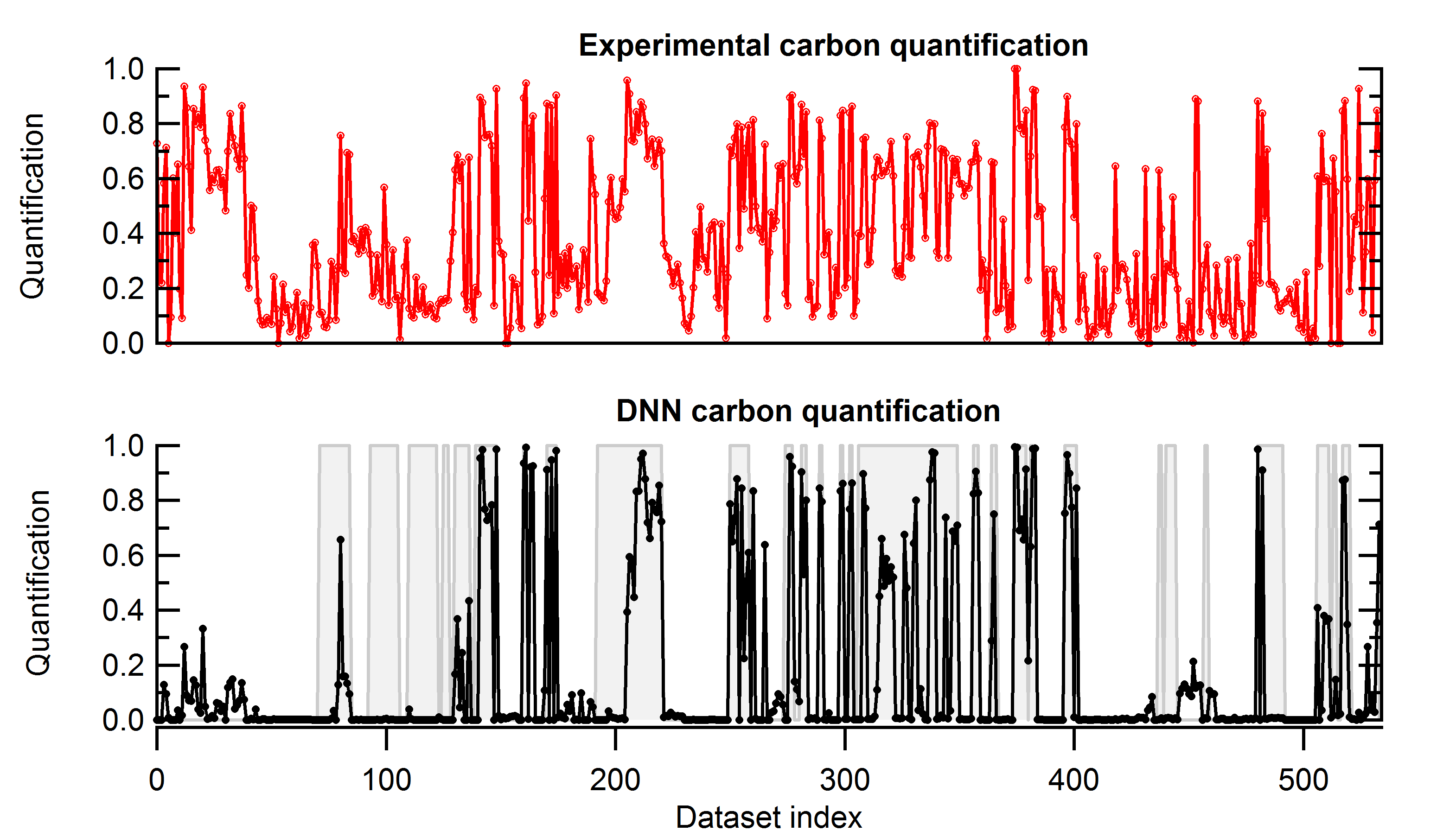}
\caption{Full carbon experimental quantification (red) and corresponding results for the DNN (black), which has been trained to ignore carbon from adventitious contamination. \label{fig_carbon}}
\end{center}
\end{figure}

Finally, we address the capability of the DNN to discriminate the carbon contamination from actual carbon-based compounds. Since it was not possible to accurately compute the level of carbon contamination of the experimental spectra, we will use a qualitative proxy measure. The red graph in Figure \ref{fig_carbon} shows the overall carbon content of the experimental dataset, evaluated from the total area of C 1s peak with respect to other elements peaks. The black trace shows the network output for the carbon quantification which, as expected, is significantly different from the experimental one. In general, the DNN performs very well reporting a high carbon content only for pristine organic materials, such as carbon nanotubes and other organic molecules, represented by the underlying light gray shades beneath the DNN identification. Some weak carbon presence is also detected by the DNN in the first 50 spectra of the experimental dataset, which are mostly composed by Ga, Se and Ge; these elements show several Auger features which are superimposed to C 1s core-level, possibly confusing even a trained human XPS user. Moreover, a thick contamination layer can also be present on top of organic compounds, further complicating the quantification process; that is, for instance, the case of the experimental spectra around the 100 dataset index.

\section{Conclusions}
In conclusion, we have shown the application of a neural network to the identification and quantification task of XPS data on the basis of a synthetic random training set. Results are encouraging, showing a detection and an accuracy comparable with standard XPS users, supporting both the training set generation algorithm and the DNN layout. This approach can easily be scaled to different photon energies, energy resolution and data range; furthermore, the DNN could be trained to provide more output values, such as the actual chemical shifts for each element, expanding the net sensitivity towards the chemical bonds classification.

\section{Acknowledgments}
G.D. gratefully acknowledge the support of NVIDIA Corporation with the donation of the Titan Xp GPU used for this research. C.M.K. acknowledges support by the Iniziativa Specifica INFN-DynSysMath. The authors thank Askery Canabarro and Marco Della Vedova for the useful discussions.

\bibliographystyle{apsrev4-1}
\bibliography{biblio}
\end{document}